\documentclass[manuscript]{aastex}
\usepackage{color}
\usepackage{natbib}
\usepackage{graphicx}

\newcommand{\degree}{^\circ}

\shortauthors{Pant et al.}
\begin{document}

\title{\bf{Twisting/Swirling Motions During a Prominence Eruption as seen from SDO/AIA}}

%% Use \author, \affil, and the \and command to format
%% author and affiliation information.
%% Note that \email has replaced the old \authoremail command
%% from AASTeX v4.0. You can use \email to mark an email address
%% anywhere in the paper, not just in the front matter.
%% As in the title, use \\ to force line breaks.

\author{ V.~Pant\altaffilmark{1}, A.~Datta\altaffilmark{2,1}, D. Banerjee\altaffilmark{1,3}, K.~Chandrashekhar\altaffilmark{4} {\sc{and}}  S. Ray\altaffilmark{5} }
\affil{$^{1}$ Indian Institute of Astrophysics, Bangalore 560~034, India.\\
        $^{2}$ HKBK College of Engineering,Bangalore 560~045, India. \\
        $^{3}$ Center of Excellence in Space Sciences, IISER Kolkata, India \\ 
        $^{4}$ Shandong Provincial Key Laboratory of Optical Astronomy and Solar-Terrestrial Environment, Institute of Space Sciences, Shandong University, Weihai, 264209 Shandong, China.\\
        $^{5}$   Department of Physics, Jadavpur University, Kolkata 700032, India}
\email{vaibhav@iiap.res.in, vaibhavpant55@gmail.com}

\begin{abstract}
A quiescent prominence was observed at north--west limb of the Sun using different channels of {\em Atmospheric Imaging Assembly} (AIA) onboard {\em Solar Dynamics Observatory} (SDO). We report and analyse twisting/swirling motions during and after the prominence eruption. %The twisting of two footpoints and a tornado like swirling motion near the footpoints of the prominence are studied. 
We segregate the observed rotational motions into small and large scale. Small scale rotational motions manifest in the barbs of the prominence while the large scale rotation manifests as the roll motion during the prominence eruption. We noticed that both footpoints of the prominence rotate in the counter--clockwise direction. We propose that similar sense of rotation in both footpoints leads to prominence eruption. The prominence erupted asymmetrically near the southern footpoint which may be due to uneven mass distribution and location of the cavity near southern footpoint. Furthermore, we study the swirling motion of the plasma along different circular paths in the cavity of the prominence after the prominence eruption. The rotational velocities of the plasma moving along different circular paths are estimated to be $\sim$ 9--40 km s$^{-1}$. These swirling motions can be explained in terms of twisted magnetic field lines in the prominence cavity. Finally, we observe the twist built up in the prominence, being carried away by the coronal mass ejection (CME) as seen in the {\it Large Angle Spectrometric Coronagraph} (LASCO) onboard {\it Solar and Heliospheric Observatory} (SOHO). 
%{\bf In this study we found that both footpoints of prominence rotate counter-clockwise which lead to the prominence eruption. To the best of our knowledge, this scenario has not been reported  earlier.} 
\end{abstract}

\keywords{Sun:Corona, Sun:Prominence, Sun:CME} 
\newpage 

\section{Introduction}

Prominences and/or filaments are ubiquitous in the solar corona. They consist of cool plasma that is embedded in the hot 
corona.   
The temperature of the prominences can range from $7500$ K to $9000$~K \citep{2010SSRv..151..243L, 2014LRSP...11....1P,2010SSRv..151..333M}. 
They are observed as a bright arcade structure off the limb, in chromospheric lines, {\it e.g.} H$\alpha$ ($6562.8$~\AA) or 
He {\sc ii} ($304$~\AA), whereas appear dark in hot coronal lines (Fe {\sc ix} $171$~\AA, 
Fe {\sc xii} $193$~\AA). On the disk they appear darker compared to 
the background due to the presence of plasma absorption processes and are known as filaments \citep{2014LRSP...11....1P}. 
 Polarimetric measurements reported the strength of magnetic field of quiescent prominences between $8$ to $10$ G for filaments 
\citep{1998ApJ...493..978L,2003ApJ...598L..67C} and for prominences \citep{2001A&A...375L..39P}.
High resolution observations in H$\alpha$ show that prominences have the finer thread like structures 
\citep[see][and references therein]{2009ApJ...704..870L,2011SSRv..158..237L}. Prominences can remain stable for a long time. Depending on different parameters like the twist in the magnetic field, 
mass loading, and ambient magnetic field they become unstable and erupt \citep{1980IAUS...91..207M,2001ApJ...552..833M,1998ApJ...502L.181A,1999ApJ...510..485A}.  

Swirling motions have been observed in the solar atmosphere. Chromospheric swirls or small scales magnetic tornadoes   
have been observed by \cite{2009A&A...507L...9W} and \cite{2012Natur.486..505W}. These authors suggested that these 
swirls provide a channel to transfer energy from the lower to the upper atmosphere. Several giant tornado like structures (large scales) 
have  also been reported in the solar atmosphere, which are associated with the vertical structure and 
barbs of  prominences/filaments \citep{2012ApJ...752L..22L,2012ApJ...756L..41S, 2013ApJ...774..123W}. However, there are only a few reports on the estimation of the rotational speed of the plasma associated with giant tornadoes \citep{2012ApJ...752L..22L}.  

A study of a tornado like prominence from {\em the extreme ultraviolet imaging Spectrometer} (EIS)  onboard Hinode and {\em Solar Dynamics Observatory/Atmospheric Imaging Assembly} (SDO/AIA) indicated a rotation of plasma around the tornado axis \citep{2014ApJ...785L...2S}.  A new model of vortex-filament was suggested by \citet{2014ApJ...785L...2S} according to which, the vortex motions twists the magnetic field. The plasma may be transported from the photosphere through this twisted magnetic field. This gives rise to highly twisted flux rope (FR) structure, which eventually becomes unstable after a large twist built up \citep{2013ApJ...774..123W,2014ApJ...785L...2S}. \citet{2014ApJ...797...52Y} studied the eruption of an active region filament. The twist build up was found to be enough for kink instability. They also confirmed that the magnetic helicity was transferred from the photosphere to the corona.
In addition to small scale twisting motions discussed above, large scale twists (roll motion) are also reported during a prominence eruption \citep[see][]{2013SoPh..287..391P}.

Furthermore, prominence eruptions are usually associated with coronal mass ejections (CMEs), which carry plasma and magnetic field into the heliosphere and are the main source of geomagnetic disturbances \citep[see ][and references therein]{2014LRSP...11....1P,2015ASSL..415..381G}.  Erupting prominences form the bright core of a typical three part CME, 
along with a bright frontal lobe, and a dark cavity \citep{2011LRSP....8....1C,1985JGR....90..275I,1981ApJ...244L.117H}.

In this article, we study small and large scale twisting and swirling motions before, during, and after a prominence eruption. We propose a new scenario for roll motion in prominence which leads to prominence eruption. A dynamical study of such motions shed new light in the understanding of the evolution of the prominence instability and eventual eruption. We observe a unique scenario of asymmetric prominence eruption and find signatures of twists built up in the prominence being carried away by the CME. Paper is organised as follows. In sections~\ref{analysis} and \ref{observations}, we describe the observation of twisting and swirling motions of the prominence as observed through different extreme ultraviolet (EUV) channels, methods of data analysis, and results. In section~\ref{prom_cme}, we describe the association of the CME with the erupting prominence, which is followed by the discussions and conclusions in section~\ref{origin and role}.

\begin{figure*}
   \centering
   \includegraphics[width=15.5cm,height=15.5cm, angle=90]{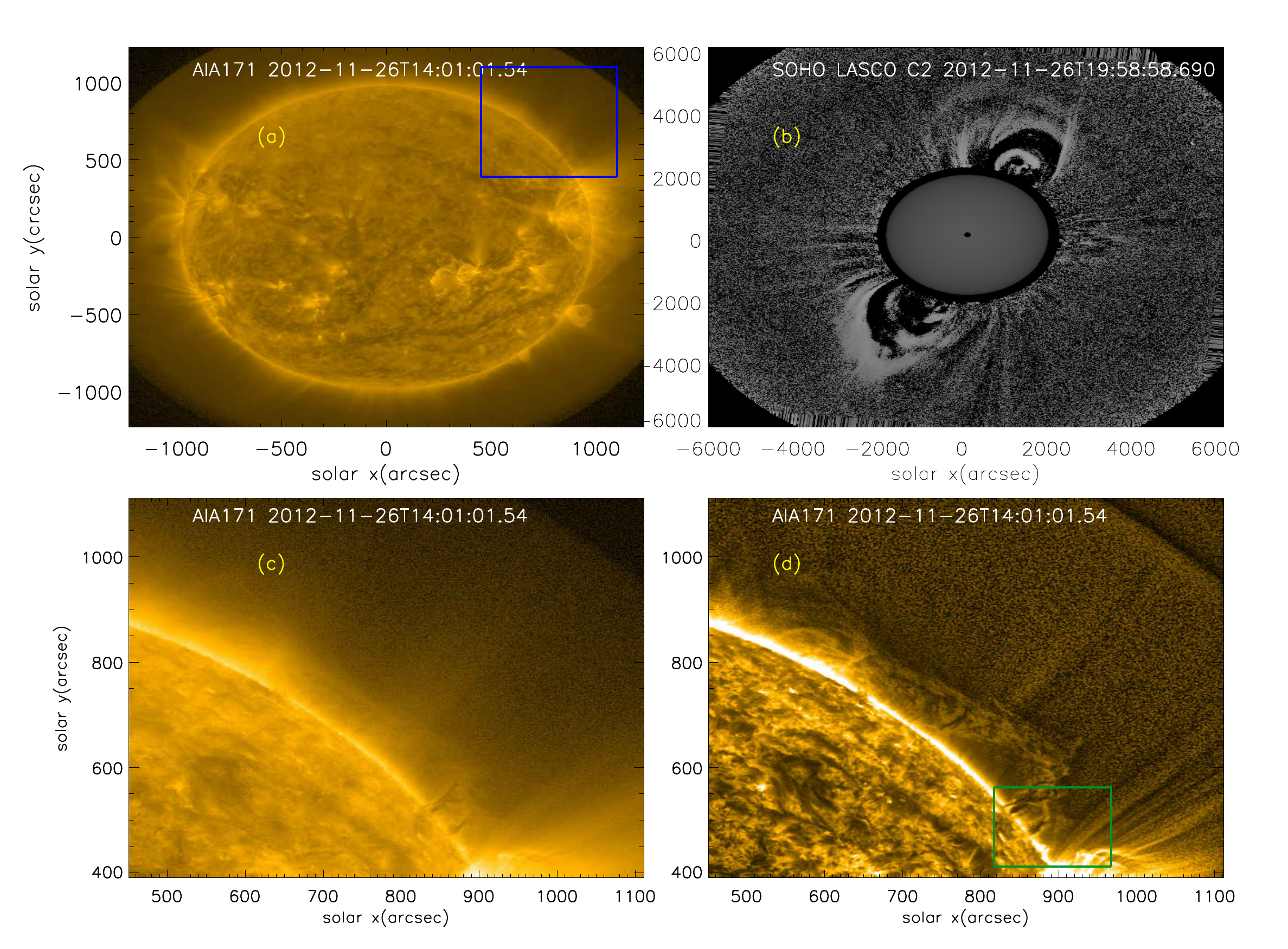}
   \caption{(a) AIA 171 {\AA} full disk image, the blue rectangle represents the Region of Interest A (ROI A). (b) A CME that is 
   associated with the prominence. An animation of this panel is available online (c) AIA 171 {\AA} image of the ROI A. (d) The ROI after applying multi-gaussian 
   normalized filter. Twisting and swirling are observed in the region marked by  green rectangle (ROI B).}
   \label{refFD}
   \end{figure*}

\section{Data and Analysis}
\label{analysis}

A prominence eruption was observed at the north-west limb of the Sun on 2012 November 26,  by 
AIA \citep{2012SoPh..275...17L} on board SDO (see, Figure \ref{refFD}a).  
 We used 304 {\AA}, 171 {\AA}, and 193 {\AA} emission images of AIA for this study. For the dynamical study, we chose a  10~h time sequence starting from 11:00 UT on 2012 November 26. % with Images of size 1100 pixels by 1200 pixels, centered at (777 arcsec, 751 arcsec). 
 The time cadence and the spatial resolution of the data are 12~s and 1.5 arcsec respectively. The standard {\it aia\_prep.pro} routine was used to process the initial level 1.0 data for correction of roll angles and to calibrate the data from different channels to a common centre and plate scale.  To focus on the twisting and swirling motions, we used one subfield with 1 hr duration. The subfield, Region of Interest (ROI) B in Figure \ref{refFD},  centered at (891 arcsec, 487 arcsec), was chosen for the study of swirling motions after the prominence eruption and is described in Section \ref{swirl}.  
This prominence eruption was associated with a CME at a central principal angle, 340$\degr$ (see Figure \ref{refFD}b), as listed in {\em the Solar and Heliospheric Observatory} (SOHO)/{\em Large Angle and Spectrometric Coronagraph} (LASCO) CME catalogue \citep{2004JGRA..109.7105Y}. The 
linear speed of the CME was 299~km s$^{-1}$. CME first appeared in the LASCO/C2 field of view at 17:48 UT. The core of the CME was rotating anti-clockwise (see animated Figure~\ref{refFD}). Such CMEs are called `cartwheel' CMEs and have been reported in  earlier studies \citep{2012SoPh..281..137K}. We discuss prominence-CME association in Section~\ref{prom_cme}. 

%\textcolor{red}{In general, a prominence consists of a horizontal spine and vertical barbs. Presence of twisted flux tubes have been reported in the horizontal part  and the vertical part is believed to be consist of many flux tubes. The magnetic twist is transferred from the photosphere to the higher corona by the rotation of these flux tubes \citep{2014ApJ...785L...2S}. The rotation and twist of these structures play an important role in the evolution of a prominence and conditions led to its eruption. On the other hand, prominence eruption, evolution of Fluxropes (FR) and coronal cavity are precursor of a CME but it is difficult to observe FRs and coronal cavity as they are quite faint.}

We further processed the AIA images to enhance faint structures at different spatial scales. First, we convolved each image with a low pass filter (Gaussian kernel) and subtracted the convolved image with the normal intensity image to retain the features that have high spatial frequency. We iterated this process three times to estimate the uncorrelated noise. Then, we subtracted uncorrelated noise from the original image. To enhance the finer structure, we used a normalized multi--Gaussian filter to the filtered image \citep[see ] []{2014SoPh..289.2945M,2015ApJ...801L...2P}. We chose the width of the Gaussian filter to be 21, 41, and 61 pixels. We added Gaussian filtered images of different spatial scales to the original image. The prominence threads and fine structures in the coronal loops were much clearly visible in the multi--Gaussian filtered images (see Figure \ref{refFD}d). Hereafter, all analysis is done using multi-Gaussian filtered images.

%\begin{figure}[ht]
 %  \centering
  % \includegraphics[width=15.5cm,height=15.5cm, angle=90]{Tornado_plotsnew/formation_FR_B.eps}
  % \caption{{\bf Overlaying loop system and the prominence as seen in AIA 171 {\AA} channel. (a) Dark funnel like structures present near the limb. (b), (c) and (d) Snapshots of different stages of formation of the background open cavity (marked with green arrows). (e) and (f)  Emergence of closed loop in the cavity (marked by two arrows in yellow).} }
 %  \label{formation}
  % \end{figure}    
   %\clearpage

\section{Observations and Results}  
\label{observations}

The prominence appeared as a dark elongated structure near the active region AR 11616, as seen in AIA 171 {\AA} emission images on 2012 November 23. Without any apparent changes, this on-disk dark elongated structure moved towards the limb. We followed the evolution of the prominence from 09:00 UT, 2012 November 25 to 22:00 UT, 2012 November 26. As the prominence started rising (09:00 UT 2012 November 26 onwards), we focussed on the various stages of the prominence eruption. In the following subsection we will describe this time evolution in a sequential manner.   

\subsection{Evolution of the prominence from 2012 November 25; 09:00 UT-- 2012 November 26; 09:00 UT:}
\label{bkgloopevolve}
 
\begin{figure*}[h]
 \centering
   \includegraphics[scale=1,angle=90]{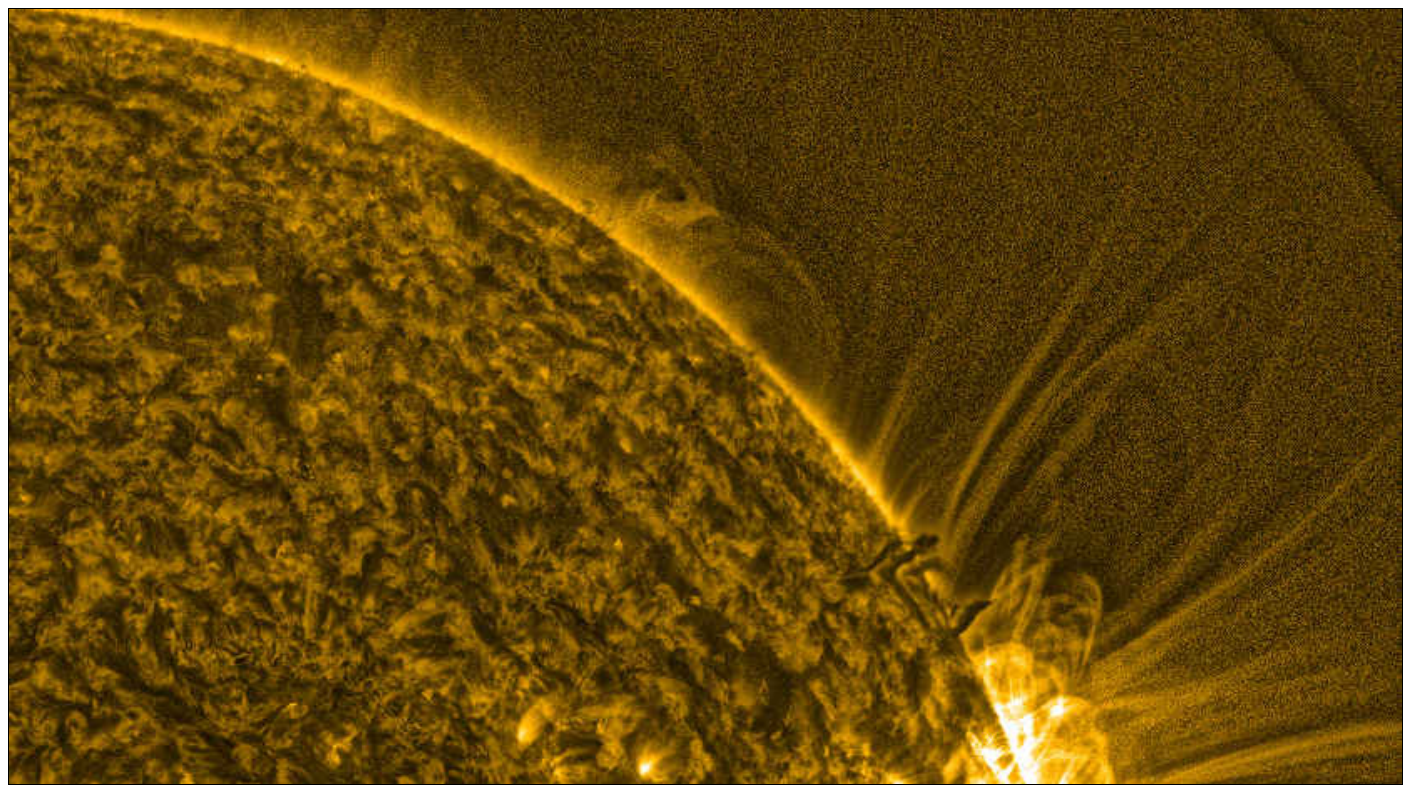}
   \caption{Movie corresponding to this animated figure shows the evolution of the prominence eruption in AIA 171~\AA~channel from 25$^{th}$ November, 2012, 09:00UT to 26$^{th}$ November, 2012, 09:00UT.}
   \label{af1}
   \end{figure*}

We note that at around 09:00 UT 2012 November 25, near the limb, the prominence barb consisted of three small dark vertical funnel like structures 
(see Figure~\ref{af1}). Overlaying loops (both closed and open) were present in the background.  As the filament barbs moved towards the limb, the spine of the filament became visible. The spine consisted of many intertwined and spiralled finer dark threads. 
 The evolution can be seen in animated Figure~\ref{af1}.
 
 %-------------
\begin{figure*}
   \centering
   \includegraphics[width=15.5cm,height=15.5cm, angle=90]{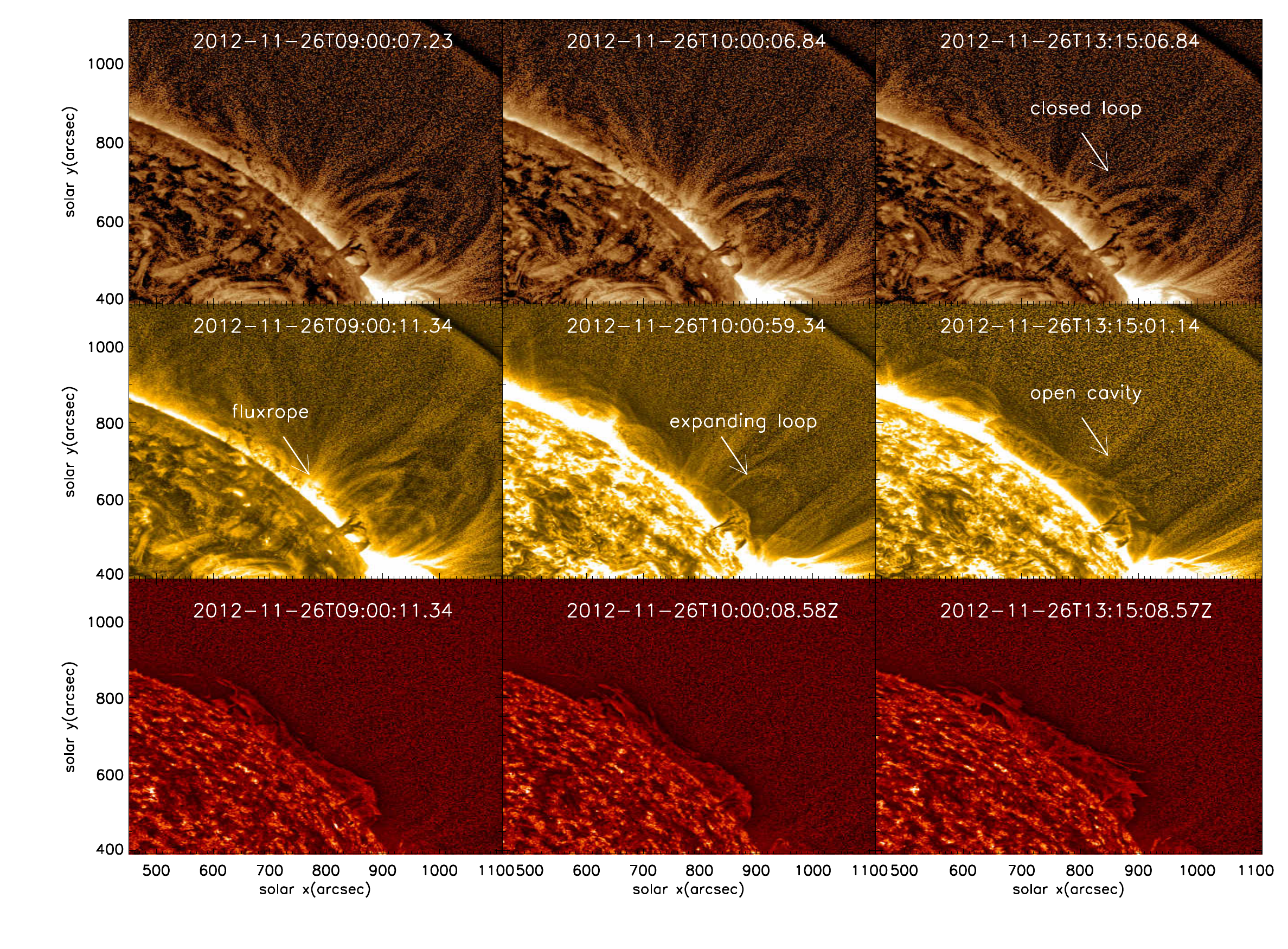}
   \caption{ (Upper panels) Prominence lift off and evolution of background loops as seen by AIA 193 {\AA} channel. 
   (middle panels) Same event observed in AIA 171 {\AA} and (lower panels) AIA 304 {\AA} channels.  }
   \label{fluxrope}
   \end{figure*} 
   
\subsection{Prominence lift off and the Eruption (2012 November 26; 09:00 UT -- 16:00 UT)}
\label{promliftoff}
%----------------

\begin{figure*}[h]
   \centering
   \includegraphics[scale=0.4]{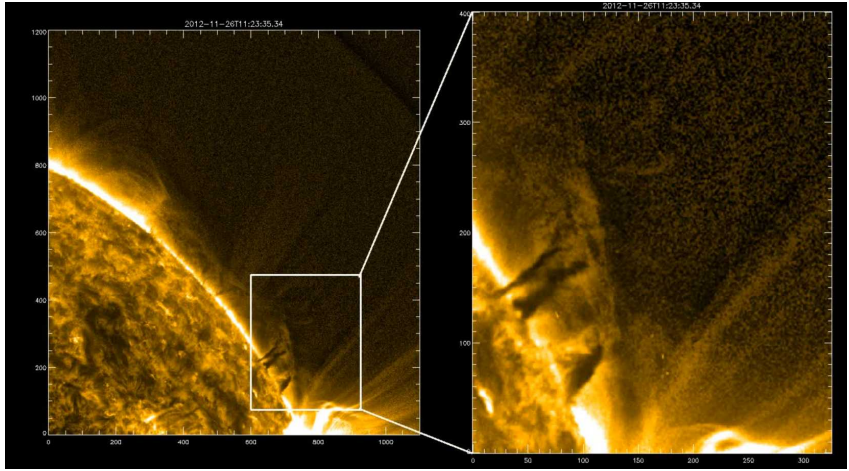}
   \caption{Movie corresponding to this animated figure shows the evolution of the prominence eruption in AIA 171~\AA~channel from 26$^{th}$ November, 2012, 09:00UT to 18:00 UT. Right panel shows the zoomed view of the vertical funnels/tornadoes.}
   \label{af5}
   \end{figure*}

After 13:00 UT 2012 November 26, the background loops seen in AIA 171~\AA~disappeared and expanded which opened the overlaying cavity (middle right panel of Figure \ref{fluxrope} and see animated Figure~\ref{af5}). As the prominence rose, it twisted and finally erupted around 16:00 UT 2012 November 26. The evolution as seen by different channels of AIA is described below.
%\begin{itemize}
%    \item {\em Evolution in AIA~304~\AA~:} 

\begin{figure*}[h]
   \centering
   \includegraphics[scale=0.4]{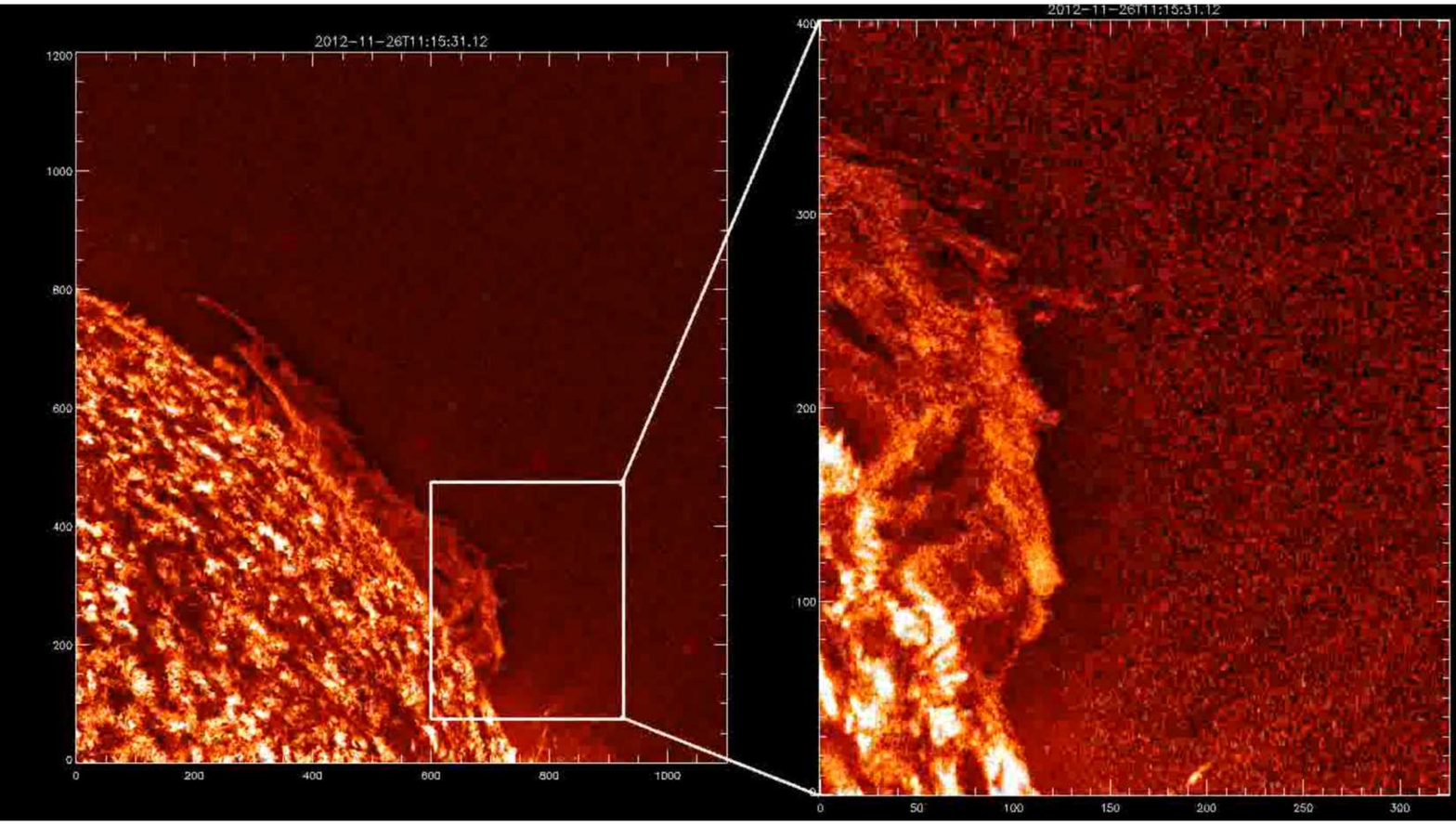}
   \caption{Same as Figure~\ref{af5} but for AIA 304~\AA~channel.}
   \label{af6}
   \end{figure*}

The prominence was visible in AIA 304 {\AA} channel as a bright structure (lower panels of Figure \ref{fluxrope} and animated Figure~\ref{af6}) that appears to consist of many dark winding threads, in hotter AIA 171 {\AA} channel (middle panels of Figure \ref{fluxrope}). Around 11:00 UT, the prominence started 
    rising slowly. The background loops as seen in AIA 171~\AA~disappeared at 13:15 UT (right middle panel of Figure \ref{fluxrope}) and subsequently the prominence started rising at a faster pace. %During the initial phase the height of the footpoints and the spine increased without any rotation or twisting motions, but 
    After 15:30 UT, during the fast rise, the entire southern footpoint started rotating around a common axis. While the northern footpoint of the prominence rose without any apparent  twist. The prominence erupted and the prominence spine broke asymmetrically near the southern footpoint around 16:00 UT.  Soon after the eruption, prominence material started falling along twisted paths. After the prominence eruption (after 16:00 UT), we observed swirling motions near southern footpoint. The entire evolution of prominence in AIA 171 and 304~\AA~ can be seen in the animated Figures~\ref{af5} and \ref{af6}, respectively.
    
The evolution of prominence in AIA~193~\AA~is same as observed in AIA~171~\AA~channel except  the evolution of the background loops (top panels of Figure~\ref{fluxrope}).  The background loops seen in AIA~171~\AA,  disappeared as prominence started rising up, but the background loops seen in AIA 193~\AA~ did not disappear but appeared to be expanded before and during the prominence eruption (right upper panel of Figure~\ref{fluxrope}).


%The rising of the cool prominence material is also observed in EUVI 304~\AA~on board STEREO A. This view shows that only a part of the prominence was erupted. 
%However, a rotating southern footpoint is clearly seen.

%%%%%%%%%%%%%%%%%%%%%%%%%%%%%%%%%%%%%%
%%%%checked upto this on 06.02.2018%%%%%%%%%%%%%%%%%%%%%%%%%%%%%%%%   
%%%%%%%%%%%%%%%%%%%%%%%%%%%%%%%%%%%%%%%%%%%%%%% 

 \subsection{Small scale twist of the prominence footpoint (2012 November 26; 11:00 UT -- 15:00 UT)}
 \label{footpoint}
In this sub--section, the small scale twisting motions are described. During the initial rising phase of the prominence, the twisting of individual dark funnels close to the southern footpoint  is 
 clearly seen in AIA 171 {\AA}. It is worth noting that the southern footpoint, where the tornado like structures and twisting motions are seen, is located near the active region while the northern footpoint is anchored at quiet region of the Sun, with no apparent twist (see animated Figure~\ref{af5}). 
 
\begin{figure}
   \centering
   \includegraphics[width=15.5cm,height=15.5cm, angle=90]{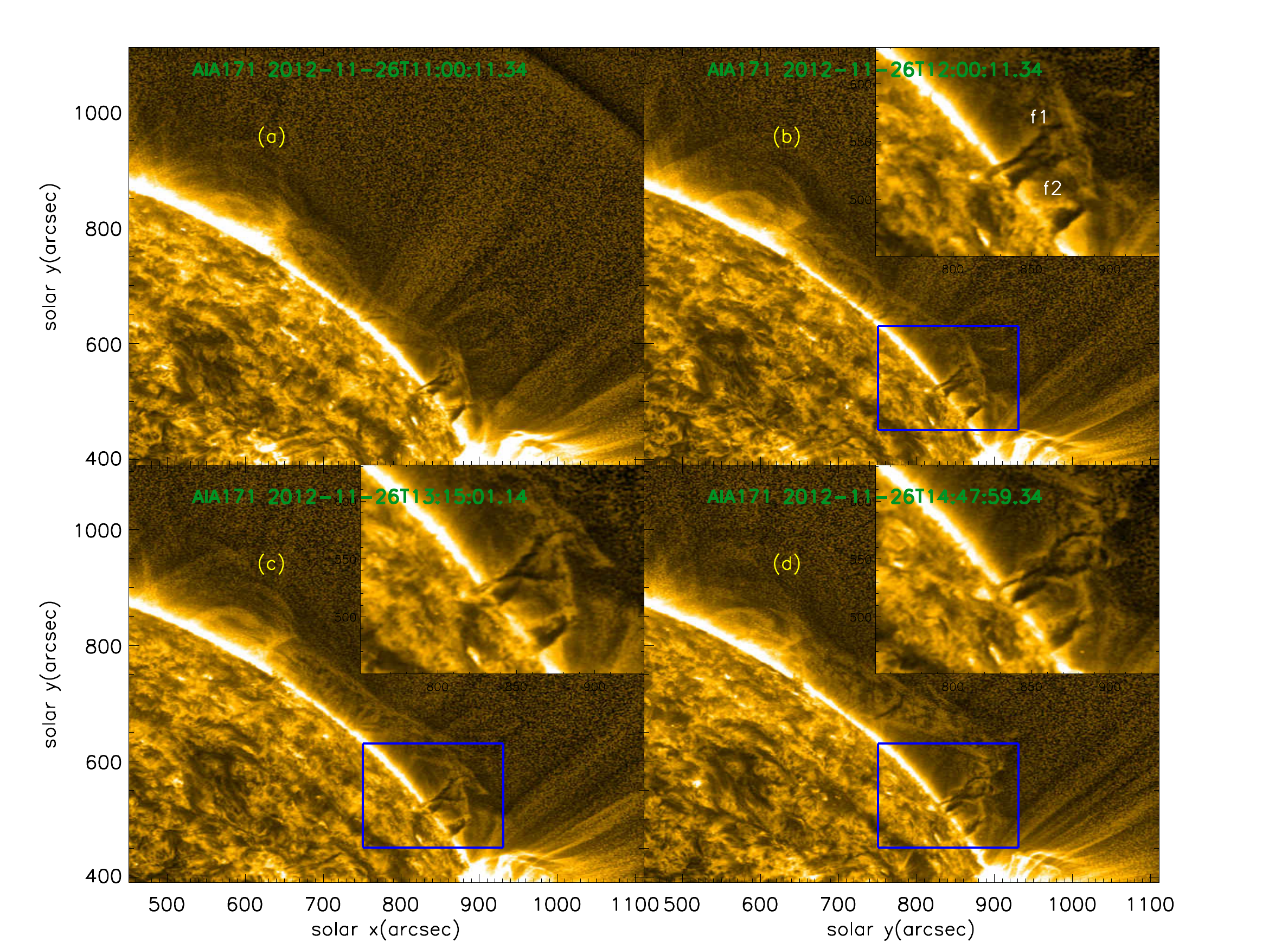}
   \caption{Rising phase of the prominence along with the twist in the southern footpoint as seen in AIA 171 {\AA} channel. The vertical dark funnels near the southern footpoint twist during the rising phase.  The blue rectangles in panels (b), (c), and (d) indicate the zoomed area shown in the corner of the corresponding panels. The dark funnels  twist, branch, and entangle, as they continue to rise. Note that the region enclosed in the blue box is different from ROI B. }
   \label{AIA171}
   \end{figure}

 Three vertical dark funnels were visible in AIA~171~\AA~around 11:00 UT at the southern footpoint (Figure~\ref{AIA171}a). During the next one hour, these dark funnels continued to rise slowly. Neither the rotation of the funnels around a certain axis nor the rotation of the funnels themselves, is clear from the images (see animated Figure~\ref{af5}). Around 12:00 UT one of the funnels (f1 in Figure~\ref{AIA171}b) divided into sub branches. The length of  the funnels along with the distance between the sub branches increased with time. Around 13:15 UT, the twist was prominent along f1 (Figure \ref{AIA171}c). This corresponds to the start of the slow rise phase of the prominence. As time progressed, the twist increased in the funnels and around 14:38 UT, apparently f1 got entangled with the adjacent funnel, f2 (see zoomed view Figure \ref{AIA171}d). Subsequently after 15:00 UT, the fast rise phase of the prominence started.
  
 \subsection {Roll motion of the spine (2012 November 26; 14:30 UT -- 17:00 UT)}
 \label{Roll}
  In this sub--section, we present the observation of the large scale twisting motion, {\it i.e,} roll motion, during the prominence eruption.
 
 As the prominence started rising after 13:00 UT, the footpoints (especially the southern footpoint) started rotating as a whole, twisting the prominence spine gradually. The roll motion is visible in AIA 304, 171, and 193~{\AA} channels (see animated Figures~\ref{af5}, \ref{af6}, and \ref{af12}).
The counter--clockwise motions of the footpoints are seen in the difference images as shown in Figure \ref{Rollfootpoint}. In the upper panels of Figure \ref{Rollfootpoint}, we follow a loop (marked by dotted line) of the southern footpoint before eruption. Since AIA 304~\AA~is an optically thick line, we can assume the observed loop in the foreground. Thus the rotation of this loop from north to south indicates that the southern footpoint rotated counter--clockwise from the top of view at the local frame. The bright loops associated with  the northern footpoint also moved from north to south as shown in the snapshots of the middle panels. Since these loops are also in the foreground, we concluded that northern footpoint of the prominence was also twisting counter--clockwise. 

After eruption, a swirling motion was observed near the southern footpoint (ROI B). We follow two plasma blobs, marked by violet and blue arrows (lower panels of Figure \ref{Rollfootpoint}), as they move along different helical paths. The blobs also show a counter--clockwise rotation of the plasma in the plane of the sky. It allows us to interpret that similar twist might have been built before the eruption.

\subsection{Asymmetric Eruption (2012 November 26; $\sim$17:00 UT)} 

\begin{figure*}[h]
   \centering
  \includegraphics[scale=0.4]{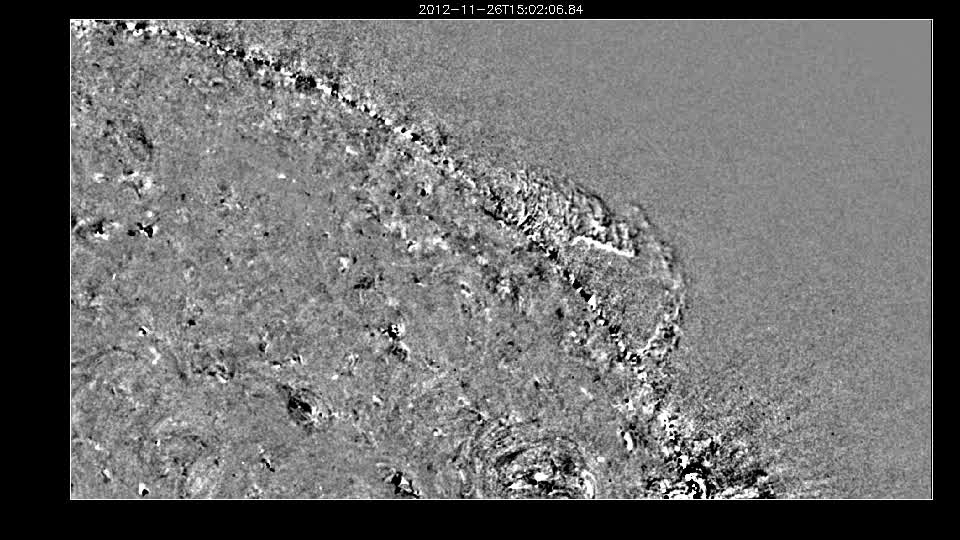}
   \caption{Movie corresponding to this animated figure represents the difference image of AIA 193~\AA. Roll motion is clearly seen in the difference image. }
  \label{af12}
  \end{figure*}

\begin{figure*}
  \centering
 \includegraphics[width=12.5cm,angle=90]{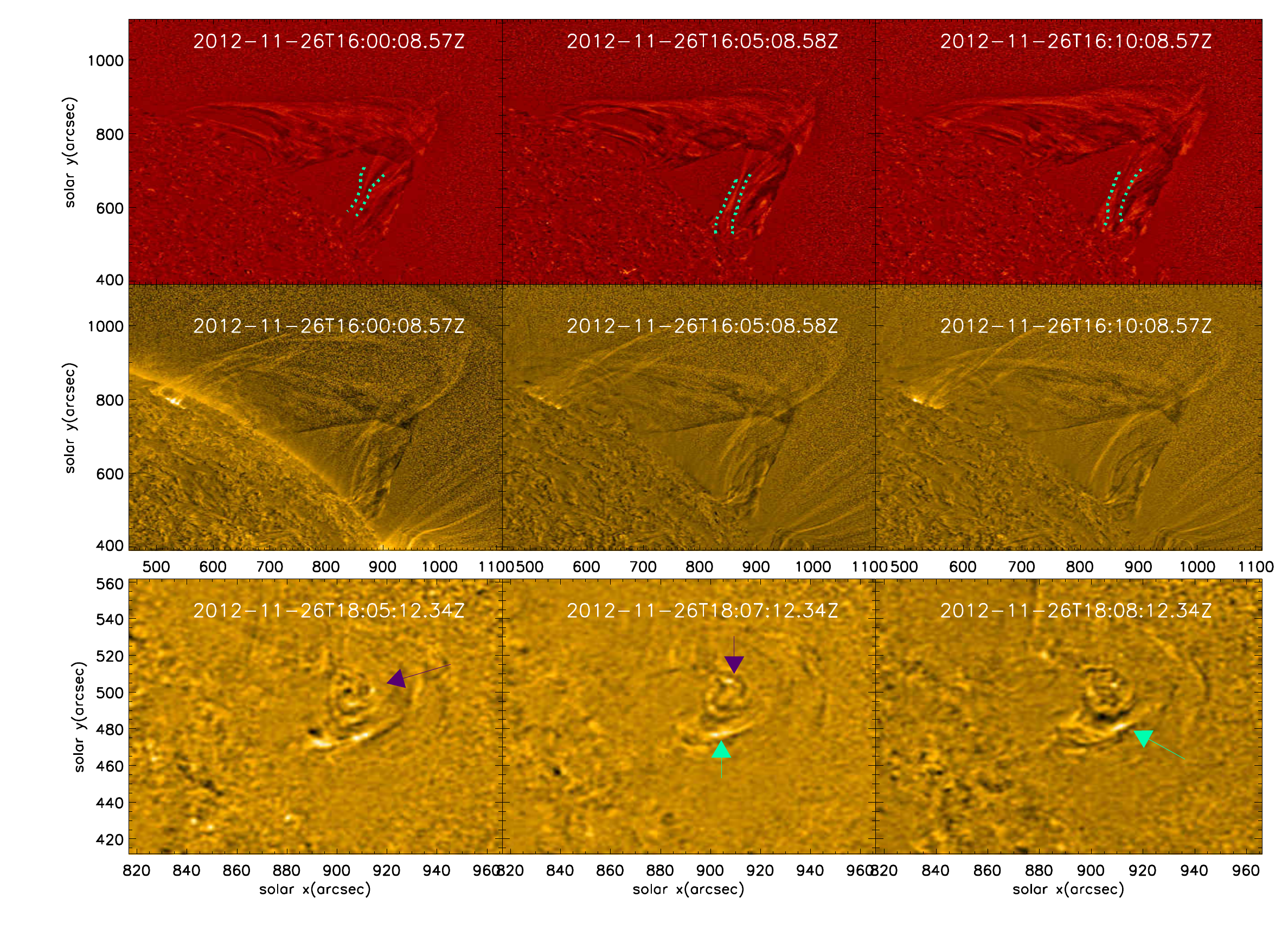}
 \caption{(upper panels) Difference images before 
 eruption in AIA 304 {\AA} channel showing anticlockwise rotation. Dashed lines show the positions of a foreground loop at southern footpoint in consecutive snapshots.
 (middle panels) Anticlockwise rotation of the northern footpoint before eruption as seen in AIA 171 {\AA} differnce 
 images. 
 (lower panels) Swirling motion of the southern footpoint after eruption in AIA 171 {\AA} channel. Tracked features are marked with 
 arrows in the consecutive snapshots.}
 \label{Rollfootpoint}
\end{figure*}

The anticlockwise twist associated with both the footpoints in turn twisted the spine and 
around 17:00 UT, the spine broke asymmetrically near the southern footpoint (see animated Figures~\ref{af5}, \ref{af6}, and \ref{af12}). We may recall that the cavity opened 
near the southern footpoint (Figures~\ref{fluxrope}) and dark funnels of cool plasma were also present near the 
southern footpoint only. The opening of the background field lines and uneven distribution of prominence material (densely distributed near southern footpoint)  might 
have caused the asymmetric eruption of the prominence near southern footpoint of the prominence.
%
%\begin{figure*}
 %  \centering
  % \includegraphics[width=13.0cm, angle=90]{Tornado_plotsnew/AIA193B.eps}
  % \caption{The prominence and background loops as seen in AIA193 channel.The {\em upper left panel} shows the cool prominence material just 
  % before it started rising slowly. In the background closed loops are clearly visible. The {\em middle left panel} shows the branching of the funnel structure near the southern (right) footpoint. The twisted spine has started rising and the closed loops are still visible. In the 
  % {\em lower left panel} we observe that the prominence has erupted and the background loops are not so prominent. The panels on the right show the difference images of 3 time frames (36 seconds). Note the helical flux tubes in the spine of the prominence.}
  % \label{AIA193}
  % \end{figure*}
 
%%%%
   %%%%% 

%%%%%%

   \begin{figure*}
   \centering
   \includegraphics[width=12.5cm, angle=90]{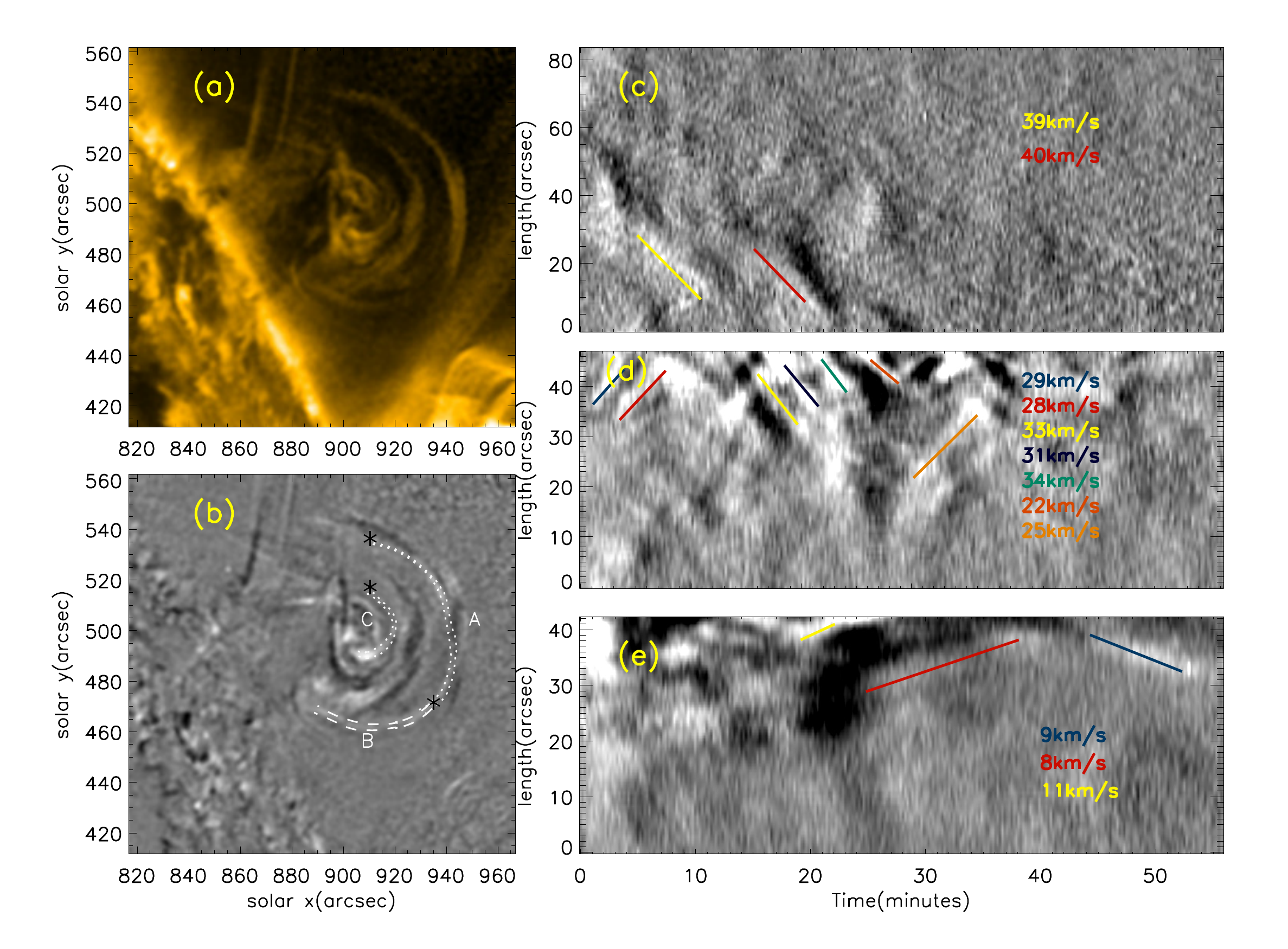}
   \caption{(a) Tornado like swirl near the right footpoint as seen in AIA 171 {\AA} channel (ROI B in 
    Figure~\ref{refFD} (d)). (b) Running difference image of the same region. Curved slits A, B, and C are used to study the 
    plasma movement during the swirl. The zero in the Y axis in the space-time maps presented in the panels (c), (d), and (e) are denoted by a ``star" symbol. (c), (d), and (e) Running difference images of space-time maps from curved slit A, B, and C 
    respectively. Speeds of plasma movement through different channels are calculated from the slope of the bright ridges. The positive and negative slopes corresponds to the clockwise and anti-clockwise rotation along the slit.}
   \label{Tornado}
   \end{figure*}
%%%%%%%%%%%%
\subsection{Swirling motion after the Eruption (2012 November 26; after 17:30 UT)}
\label{swirl}
Subsequently, around 17:30 UT, a swirling motion started close to the southern footpoint (ROI B in Figure\ref{refFD}d). 
The motion is best seen in AIA 171 {\AA} channel (see right hand panel of animated Figures~\ref{af5}). 
We observed clockwise, as well as anti-clockwise movement on the plane of the sky of the plasma along different circular paths  (Figure~\ref{Tornado} (a)). We chose two curved slits, namely A and B along outer channels, 
and another curved slit, C, closer to the axis of swirling structure as shown in Figure~\ref{Tornado} (b) to study the swirls. 
A 4 minute running difference images of the space-time maps along these channels exhibit bright and dark ridges (Figures~\ref{Tornado} (c), (d), and (e)). Bright ridges with negative (positive) slope indicate  plasma movement in anti-clockwise (clockwise) direction. The plasma speeds are estimated from the slopes of the bright ridges of the space-time maps. Corresponding to the slit location A we observed  anti-clockwise motions of plasma with speed $\sim$ $35-60$ km~$s^{-1}$, 
whereas for B and C locations both anti-clockwise and clockwise movements were recorded. %Plasma swirls with speeds $\sim$ $35-60$ km~$s^{-1}$ along A. The speeds are found to be $22-29$  km~$s^{-1}$ along `B'. Speeds along the inner curved slit, `C' are around $11-12$ km~$s^{-1}$, which are much lower compared to that along the outer curved slits. We observe that speeds  do not vary for two different directions of motion. 
 It appears that the plasma follows the helical or twisted field lines which looks like swirling motion. At a later time,  the magnetic field lines relaxed and spiral structure was no longer present.  
%
%%%% corrected upto this on 26.12.2017%%%%%%%%%%%%%%%%%%
%%%%%%%%%%%%corrections done%%%%%%%%%%%%%%%
%Vortex flows are observed in the photosphere. As the plasma sinks down, bathtub effect is observed due to conservation of momentum. If the footpoint of a magnetic structure in upper layer coincides with such vortex flow below, then another type of vortex motion can develop \citep{2013ApJ...774..123W,2014PASJ...66S..10W}. We have observed that plasma speed is greater in the outer channel. This is a consequence of higher centrifugal force along the outer channel. Plasma can be accelerated upwards and these swirls provide a mechanism of energy supply from lower to upper atmosphere \citep{2012Natur.486..505W}.   

\section{Prominence eruption and  Onset of ``Cartwheel" CME}
\label{prom_cme}
\begin{figure*}
  \centering
 \includegraphics[scale=0.5]{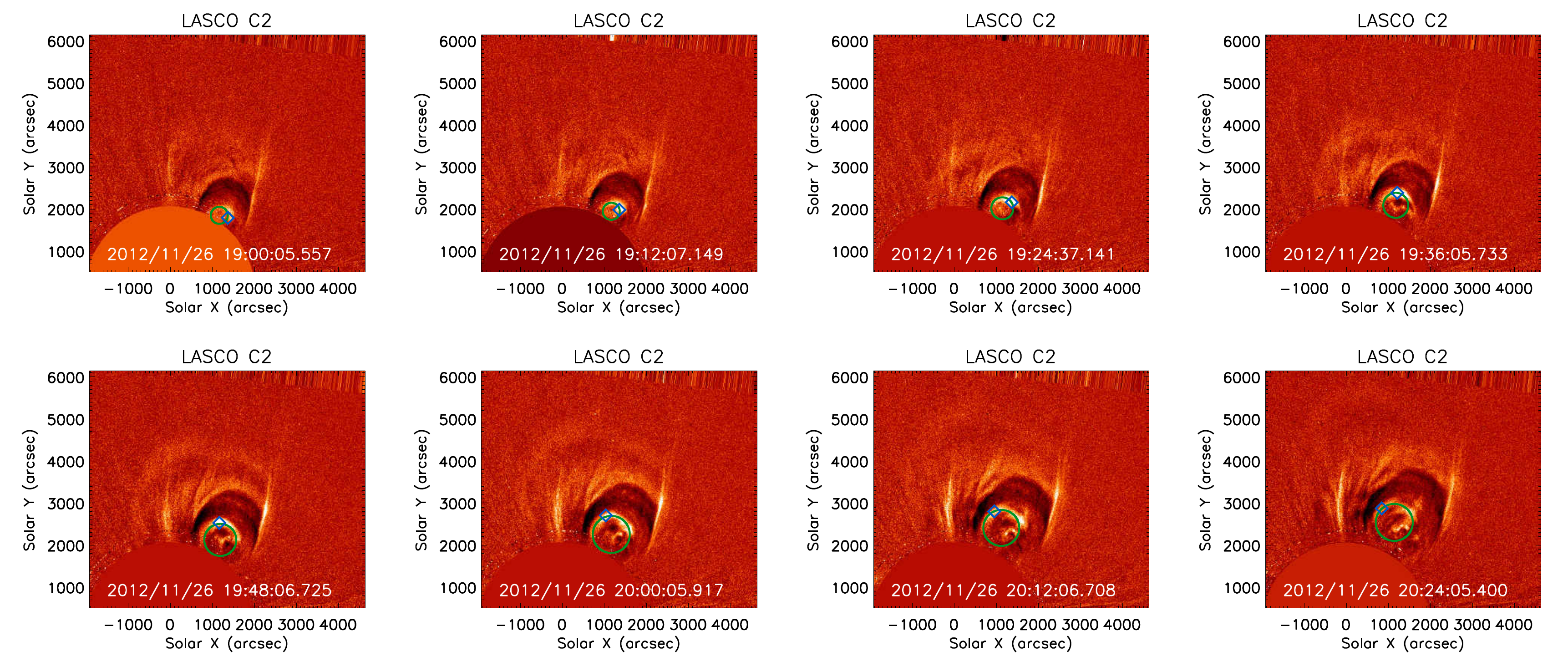}
 \caption{LASCO/C2 running difference image. Diamond in blue represents the feature that is rotating anticlockwise. Since the feature is extended, diamond symbol represents the location of maximum brightness. Circle in green passes through the feature of interest and is used to estimate the angular position of the feature. The degree of rotation is 150$^{\degree}$. }
 \label{cme}
\end{figure*}

\begin{figure*}
  \centering
 \includegraphics[scale=0.5]{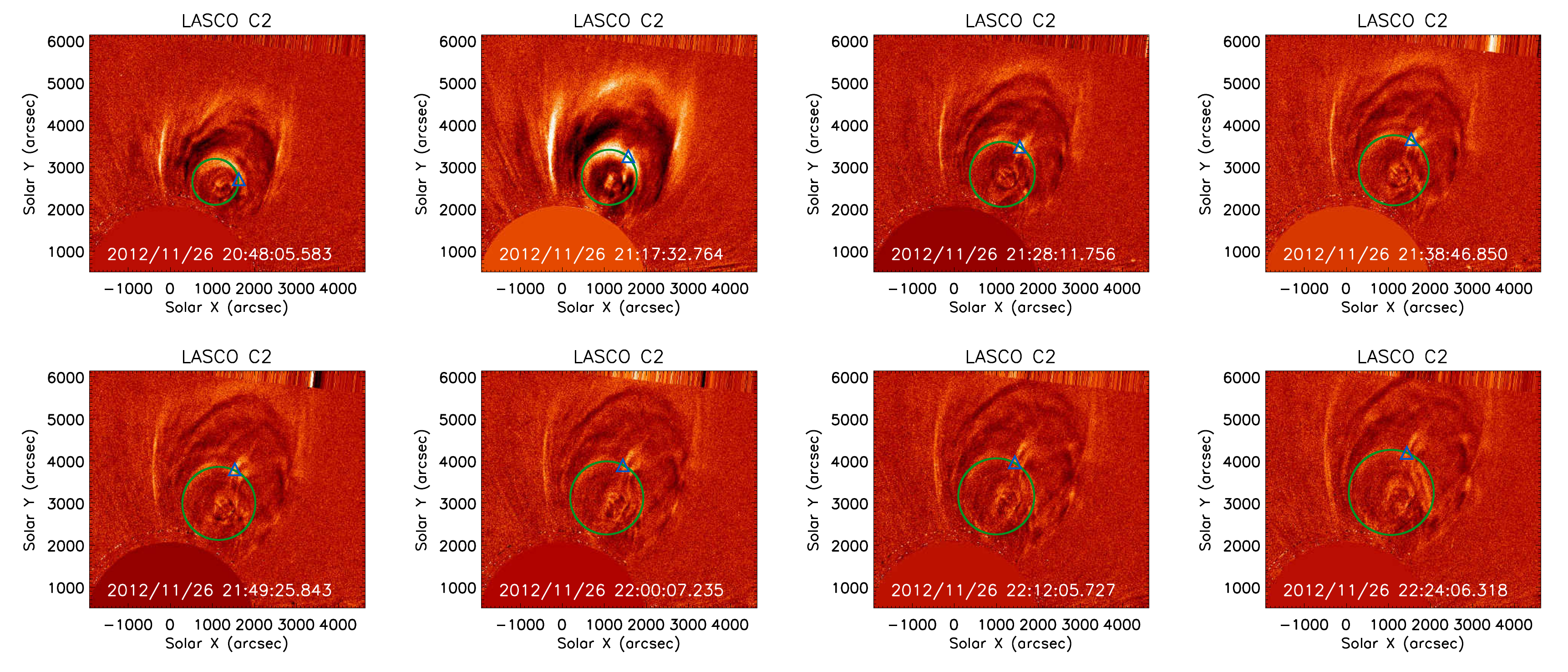}
 \caption{Same as Figure~\ref{cme} except for a different feature marked with a blue triangle. Degree of rotation is 64$^{\degree}$.}
 \label{cme1}
\end{figure*}

In this section we will study the properties of the CME  associated with the prominence eruption. Although most of the prominence material fell back, a part of it was carried away by the CME. We observe that the twist which was built up around the southern footpoint of the prominence, manifested as counterclockwise rotation (see animated Figure~\ref{refFD}) of the core of the CME. It should be noted that due to the spatial gap between the field of views of AIA and LASCO, a spatial relationship between twist in prominence and the rotation of the CME could not be established. The outward propagation of the CME along with the rotation was difficult to be captured by the low cadence (12 minutes) of  LASCO/C2.   Thus, we try to estimate the degree of rotation of the features in the CME by tracking some of the prominent features. In order to estimate the degree of rotation, first we choose a feature which was found to be rotating clearly in the coronagraph images. Then the angular position of the feature was calculated. To estimate the angular position of the feature, a circle is fitted such that it passes through the points selected from the bright feature. Angular position is the angle that the feature makes with the horizontal direction. In subsequent images, as the CME moved outward, we estimate the angular position of the feature at every frame. Therefore, as the feature rotated, the angular position changed. We track the feature until it either disappeared from the coronagraph images or became too faint to be detected. We apply this procedure to two features as identified in the coronagraph images (see Figures~\ref{cme} and \ref{cme1}) and note that the they rotated by 150$^{\degree}$ and 64$^{\degree}$, respectively. First feature appeared at 19:00 UT and disappeared after 20:24 UT while the second feature appeared at 20:48 UT. Therefore, the total degree of rotation is the sum of the degree of rotation of both features, {\it i.e,} 214$\degree$ assuming that these two features were the part of same prominence. It should be noted that the degree of rotation will not be affected by the projection in the plane of sky. This is still an underestimate of the total twist because we couldn't see clear features at all frames.
%\begin{figure*}[h]
%   \centering
%   \includegraphics[scale=0.5]{Tornado_plotsnew/movie14.png}
%   \caption{Movie corresponding to this animated figure shows the CME in LASCO/C2 FOV at the location of the prominence eruption. Features inside the CME are seen clearly to be rotating anticlockwise.}  \label{af14}
%   \end{figure*}  

\section{Discussions and Conclusions}
\label{origin and role} 
In this article we study a dynamical evolution of a quiescent prominence eruption which includes twisting, swirling motions, and an association with a CME. Here we summarise the different stages of the evolution of the prominence eruption and salient features associated with it.

We found that the overlying loops as seen in AIA 171~\AA~ disappeared prior to the prominence lift-off and as expected, subsequently the prominence erupted.  However, neither the signature of the FR nor the background loops were seen in hotter channels of AIA, 94 and 335~\AA.
%\item {\it Small scale twisting motions}: 
We considered barbs as rotating funnel like structures. Since the length scale of the rotation in barbs is small as compared to the roll motion of the prominence, we termed the rotation of barbs as small scale twisting motion. Such rotating structures have been termed as giant tornadoes by \cite{2013ApJ...774..123W}. The southern footpoint of the prominence was lying in close proximity to the active region. Photospheric vortex flows are more frequent near the active region \citep{2012A&A...538A..62B,2014SoPh..289.4481D} thus it may explain why the southern footpoint had manifested larger twist as compared to the northern footpoint which was anchored at the quiet region of the Sun. However, it should be noted that the photospheric vortex motions cannot be measured because the footpoints of the prominence were at the limb. Furthermore, the rotation of the tornado could transfer cool material and magnetic twist to upper layers. The magnetic twist increased the coronal magnetic field stress and the prominence became unstable. This scenario is consistent with \citet{2014ApJ...785L...2S}.

\begin{figure*}[h]
  \centering
 \includegraphics[scale=0.5]{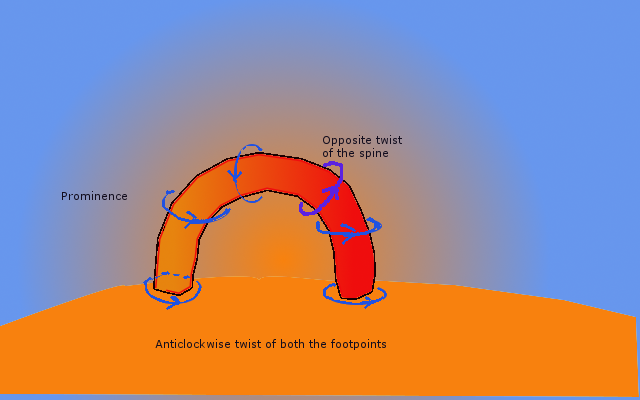}
 \caption{A cartoon showing the twist at both footpoints of the prominence in the same direction. When the twist propagates upward, it twists two sides of the spine of the prominence in opposite directions which cause prominence spine to break. }
 \label{cartoon}
\end{figure*}

In contrast to \citet{2013SoPh..287..391P}, we observed that twist in both footpoints of the prominence was anticlockwise. We describe this scenario with a cartoon as shown in Figure~\ref{cartoon}; where we have shown that the anticlockwise twist of the two footpoints results in the opposite twist at the spine of the prominence which leads to the prominence eruption. This is the main finding from our study. To our knowledge, this scenario has not been reported earlier.
%\item {\it Asymmetric eruption of the prominence}: 
We noted that the prominence erupted asymmetrically near the southern footpoint. Several factors may contribute to the asymmetric eruption. First, we found the signatures of expanding cavity above the prominence in AIA 171 and 193~\AA~ images. From Figure~\ \ref{fluxrope}, we noted that the magnetic field configuration was different over two footpoints of the prominence. The cavity was located near the southern footpoint and we observed that the prominence started rising at the location of the open cavity. We have not observed any signature of possible reconnection during this event. So the reason behind the change in the magnetic topology is not clear to us. Second, we believe that the prominence material was densely distributed at the southern footpoint due to the presence of dark funnel like structures.  Uneven distribution of the prominence material and the open cavity located near the southern footpoint facilitated the easy lift-off of the prominence near the southern footpoint. This scenario of the asymmetric eruption of the prominence is also a new result that has not been reported earlier.

We noted the swirling motion of the plasma along several circular paths after the prominence eruption. It appears that the swirling motion was caused by the falling material along the helical field lines that were located around the dark funnels/tornadoes. This is consistent with the numerical model of  \citet{2014ApJ...792L..38X}. The authors have simulated the condensation process in a solar prominence, with  the helical field lines in the cavity of the prominence \cite[see Figure~5 of ][]{2014ApJ...792L..38X}.  After eruption the plasma blobs may follow these helical field lines and cause swirling motions.  We conjecture that the helical field lines might have resulted from  the twist that was built up around the southern footpoint of the prominence. Finally, these helical field lines might have disappeared and the shape might have completely changed, which can explain the disappearance of the dark funnels/tornadoes and spiral structure. Furthermore, we found that a part of the twist was also carried away by the CME associated with the prominence eruption.  We estimated the degree of rotation $\sim$ 214$\degree$ at $\sim$ $3$ solar radii. Such a high degree of rotation has not been reported in earlier studies. A similar event, nicknamed as `cartwheel CME' was observed by \cite{2012SoPh..276..241T,2011A&A...525A..27P}. A rotation of $115^{\degree}$ was estimated at $2.5$ solar radii which is much less as compared to the degree of rotation reported in this study.\\
The study of the rotation is important to understand the eruption mechanism of the FR \citep{2012SoPh..281..137K}. For instance, the rotation of the FR and the associated CME may indicate the occurrence of helical Kink Instability \citep[KI;][]{2012SoPh..281..137K}. \citet{2012SoPh..281..137K} performed numerical simulations of a `cartwheel' CME that resulted from an erupting FR. In the numerical model, they used external toroidal field component pointing along the FR. They proposed that the rotation by KI depends on the initial twist in the FR, the strength, and the height profile of the overlying field. If overlying field decreases slowly with height then the rotation will be strong at small heights. Otherwise, the rotation will be distributed over large heights. They tried to fit the observations by varying the strength of the external toroidal field (sheer field) and the initial twist in the FR. Thus the degree of rotation provides a strong constraints for the numerical modeling. The authors also proposed that twist in a FR (5$\pi$) alone is not enough to match the rotational profile. The inclusion of the external sheer field is required to match the observed rotation.
Thus the degree of rotation of a CME may be used to probe the magnetic field topology of the overlying field and its variation with height by performing numerical modeling similar to \citet{2012SoPh..281..137K}. The high degree of rotation, observed in this study, may provide strong constraints on the existing models. We conjecture that the high degree of rotation may happen due to highly sheared external magnetic field.  

Thus we conclude that the small scale twist of dark funnels/tornadoes present near the southern footpoint might be responsible for the large scale twist in the footpoints of the prominence (roll motion) during the prominence eruption. Similar sense of rotation in both footpoints of prominence caused the prominence spine to break which lead to the prominence eruption. The anti-clockwise rotation at the footpoints of the prominence manifested as rotating features inside the CME. Thus we feel that this morphological study of prominence eruption may provide a new insight into the prominence eruption models. However, further studies with better quantitative estimates are needed to confirm some of the findings as reported here.
%

%%%%%%%%%%%%%%%%%%%%%%%%%%%%%%%%%%%%%%%%%%%%%%%%
%%%%%%%%%%%%%%%%%%%%%%%%%%%%%%%%%%%%%%%%%%%%%%%%

\acknowledgments
Authors thank anonymous referee for his/her valuable comments that has helped us to substantially improve the manuscript. The authors would also like to thank Yingna Su for her constructive comments. AIA/SDO data are provided by the Joint Science Operations Center Science Data Processing. The SDO Data used is the courtesy of NASA/SDO and the AIA, EVE, and HMI science teams.

\clearpage
%\bibliographystyle{aasjournal}
%\bibliography{prominence}

\end{document}